\documentclass{svmult}
\usepackage{graphicx}
\usepackage{subfigure}
\usepackage{amssymb}
\usepackage{amsmath}
\usepackage{stmaryrd}
\usepackage{tikz}

\newcommand{\G}{\mathcal{G}}
\newcommand{\J}{\mathcal{J}}
\newcommand{\T}{\mathbb{T}}

\newcommand{\topo}[1]{{#1}}

\newcommand{\TVG}{\ensuremath{\G=(V,E,\T,\rho,\zeta)}}
\newcommand{\TVGL}{\ensuremath{\G=(V,E,\T,\rho)}}

\usepackage{makeidx}     
\usepackage{graphicx}    
\usepackage{multicol}    

\makeindex             

\begin{document}

\title*{Emergence through Selection: The Evolution of a Scientific Challenge}
\author{Walter Quattrociocchi\inst{1}\and
Frederic Amblard\inst{2}}
\institute{University of Siena
Pian Dei Mantellini 44, 53100, Siena, Italy\\
Tel: +39 0577 233 710\\
\texttt{walter.quattrociocchi@unisi.it}
\and IRIT - Université Toulouse 1 Capitole,
2, rue du Doyen Gabriel Marty\\
31042 Toulouse Cedex 9\\
Tel: +33 (0) 561 128 795\\
\texttt{frederic.amblard@univ-tlse1.fr}}
%
%
\maketitle

\begin{abstract}
One of the most interesting scientific challenges nowadays deals with the analysis and the understanding of complex networks' dynamics and how their processes lead to emergence according to the interactions among their components. 

In this paper we approach the definition of new methodologies for the visualization and the exploration of the dynamics at play in real dynamic social networks.  We present a recently introduced formalism called TVG (for time-varying graphs), which was initially developed to model and analyze highly-dynamic and infrastructure-less communication networks.  As an application context, we chose the case of scientific communities by analyzing a portion of the ArXiv repository (ten years of publications in physics). The analysis presented in the paper passes through different data transformations aimed at providing different perspectives on the scientific community and its evolutions. 

On a first level we discuss the dataset by means of both a static and temporal analysis of citations and co-authorships networks. Afterward, as we consider that scientific communities are at the same time communities of practice (through co-authorship) and that a citation represents a deliberative selection pointing out the relevance of a work in its scientific domain, we introduce a new transformation aimed at capturing the interdependencies between collaborations' patterns and citations' effects and how they make evolve a goal oriented systems as Science. 

Finally, we show how through the TVG formalism and derived indicators, it is possible to capture the interactions patterns behind the emergence (selection) of a sub-community among others, as a goal-driven preferential attachment toward a set of authors among which there are some key scientists (Nobel prizes) acting as attractors on the community. 
\end{abstract}



\section{Introduction}

One of the most interesting scientific challenges nowadays deals with the analysis and the understanding of social networks' dynamics and how their processes lead to emergence according to the interactions at play among their components.
The research efforts in this area strive to understand what are the driving forces behind the evolution of social networks and how they are articulated together with social dynamics, e.g., opinion dynamics, the epidemic or innovation diffusion, the teams formation and so on  (\cite{amblard01,Moore2000,Lelarge09,Carley02,Powell05,Guimera05,QuattrociocchiPC09,castellano07,quattrociocchi2010e}). In this paper we approach the definition of new methodologies for the visualization and the exploration of the dynamics within real dynamic social networks. 
As an example, we chose the case of scientific communities by analyzing a portion of the ArXiv repository (ten years of publications in physics) focusing on the social determinants (e.g. goals and potential interactions among individuals) behind the emergence and the resilience of scientific communities. 
In particular, the analysis addresses the co-existence of co-authorships' and citations' behaviors of scientists by focusing on the most proficient and cited authors interactions' patterns and, in turn, on how they are affected by the selection process of citations. 
Such a “social” selection a) produces self-organization because it is played by a group of individuals which act, compete and collaborate in a common environment in order to advance Science and b) determines the success (emergence) of both topics and scientists working on them. 
 
On the one hand, the studies on scientific network dynamics deal with the understanding of the factors that play a significant role in their evolution, not all of them being neither objective nor rational – e.g., the existence of a star system \cite{Wagner2005}, \cite{Newman2001}, \cite{Newman2004}, \cite{Barabasi2002} the blind imitation concerning the citations \cite{MacRoberts96}, the reputation and community affiliation bias \cite{Gilbert77}. On the other hand, having some elements to understand such dynamics could enable for a better detection of the hot topics and of the vivid subfields and how the scientific production is advanced with respect to selection process inside the community itself.  Among the available data to analyze such a system, a subset of the publications in a given field is the most frequently used such as in \cite{Solla1965}, \cite{Newman2001a}, \cite{Newman2004a}, and \cite{Radicchi2009}. The scientific publications correspond to the production of such a system and clearly identify who are the producers (the authors), which institution they belong to (the affiliation), which funded project they are working on (the acknowledgement) and what are the related publications (the citations), having most of the time a public access to these data explain also a part of its frequent use in the analyses of the scientific field. 
Classical analyses concern either the co-authorships network (\cite{Barabasi2002, Newman2001}) or the citation network (\cite{Hummon89, Redner05}), more rarely the institutional network (\cite{Powell05}). Moreover, such networks are often considered as static and their structure is rarely analyzed overtime (an exception is the one performed by \cite{Radicchi2009} on Physical Review). 

The illustrative analysis presented in the paper passes through different data transformations aimed at providing different perspectives on the scientific network and its evolutions.  
On a first level we discuss the dataset by means of both static and temporal analysis of citations and co-authorships networks.
A second level of analysis consists in transforming the data in order to explicit the interdependencies between the co-authorships and citations by analyzing the scientists' representations of the collaboration structure within the scientific field. Such a representation is captured through the network of cited collaborations (\cite{QA2010a}), i.e. from a publication we have several references to other papers, each one corresponds to a promotion of the scientists authoring the work.

One of the problem when trying to characterize such a dynamic structure is that classical indicators from either graph theory or social network analysis cannot be applied directly. Therefore, we used an algebra, the Time-Varying Graphs (TVG) (\cite{CFQS2010}) that enables to take into account the dynamical aspects of networks and allows for the definition of temporal indicators (\cite{ACFQS2010a}) to characterize patterns in evolving structures. 

Through our approach, we capture the attractiveness played by famous authors on co-authorship behaviors and on the sub-communities structural evolution.

\section{Context}
In \cite{Newman2001} the network of scientific collaborations, explored upon several databases, shows a clustered and small world structure. Moreover, several differences between the collaborations' patterns of the different fields studied are captured. Such differences have been deepened in \cite{Newman2004} with respect to the number of papers produced by a given group of authors, the number of collaborations and the topological distances between scientists. 
Peltomaki and Alava in \cite{peltomaki2006} propose a new emulative model aimed at  approximating the growth of scientific networks, by incorporating bipartition and sub-linear preferential attachment. 
A model for the self-assembly of creative teams based on three parameters (e.g. team size, the rate of newcomers in the scientific production and the tendency of authors to collaborate with the same group) has been outlined in \cite{Guimera05}. 
Connectivity patterns in a citations network have been studied in relation to the development of the DNA theory \cite{Hummon89}. 
The work of Klemm and Eguiluz ( \cite{Klemm02}) observed that real networks (e.g. movie actors, co-authorship in science, and word synonyms) growing patterns are characterized by a clustering trend that reaches an asymptotic value larger than regular lattices of the same average connectivity.

In the field of social network analysis several works have approached the problem of temporal metrics \cite{Holme05, Kostakos09, KosKW08}. 
Actually, the focus is on the definition of instruments able to capture the intrinsic properties of complex systems' evolution, that is, characterizing the interdependencies and the co-existence between local behaviors (interactions) and their global effects (emergence)  \cite{Davidsen2002,Mataric92,Woolley1994,amblard01,quattrociocchi2010e}. 
The research approach to characterize the evolution patterns of social networks, at the very beginning was mainly based upon simulations, while in the past few years, due to the large availability of real datasets, either the methodology of analysis and the object of research have changed (\cite{Roth10b,LES07,KosKW08,castellano07,LESK10}). 
In particular, in the latter paper Leskovec states as central problem, for the social networks in general and for the scientific communities networks analysis in particular, the definition of mathematical models able to deal and to reproduce all the properties of dynamical real world networks such as the shrinking diameter (\cite{Les05}), or the ``small world'' effect \cite{WATTS99}. Actually instruments and paradigms affording this challenge are mainly based upon stochastic definitions (\cite{Les05b}) or conceptualized as a sequence of snapshots of the network at different times \cite{TSM+09}.

\section{Preliminaries}

\subsection{Time-Varying Graphs}
The {\em time-varying graph} (TVG) formalism, recently introduced in~\cite{CFQS2010}, is a graph formalisms based on an {\em interaction-centric} point of view and offers concise and elegant formulation of temporal concepts and properties \cite{ACFQS2010a}.

Let us consider a set of entities $V$ (or {\em nodes}), a set of relations $E$ among entities ({\em edges}), and an alphabet $L$ labeling any property of a relation ({\em label}); that is, $E \subseteq V \times V \times L$. 
The set $E$ enables multiple relations between any given pair of entities, as long as these relations have different properties, that is, for any $e_1=(x_1,y_1,\lambda_1)\in E,e_2=(x_2,y_2,\lambda_2) \in E$, $(x_1=x_2 \wedge y_1=y_2 \wedge \lambda_1=\lambda_2)\implies e_1=e_2$.

Relationships between entities are assumed to occur over a time span $\T \subseteq \mathbb{T}$, namely the {\em lifetime} of the system. 
The temporal domain $\mathbb{T}$ is assumed to be $\mathbb{N}$ for discrete-time systems or $\mathbb{R}$ for continuous-time systems. 
The time-varying graph structure is denoted by the set \TVG, where
$\rho: E \times \T \rightarrow \{0,1\}$, called {\em presence function}, indicates whether a given edge is present at a given time, and
$\zeta: E \times \T \rightarrow \mathbb{T}$, called {\em latency function}, indicates the time it takes to cross a given edge if starting at a given date. As in this paper the focus is on the temporal and structural analysis of a social network, we will deliberately omit the latency function and consider TVGs described as \TVGL.

\subsubsection{TVGs as a sequence of footprints.}
\label{sec:foot}
Given a TVG \TVGL, one can define the {\em footprint} of this graph from $t_1$ to $t_2$ as the static graph $G^{[t_1,t_2)}=(V,E^{[t_1,t_2)})$ such that $\forall e \in E, e \in E^{[t_1,t_2)} \iff \exists t \in [t_1,t_2), \rho(e,t)=1$. In other words, the footprint aggregates interactions over a given time window into static graphs. Let the lifetime  $\T$ of the time-varying graph be partitioned in consecutive sub-intervals $\tau = [t_0,t_1), [t_1,t_2) \ldots [t_i,t_{i+1}), \ldots$; where each $[t_k,t_{k+1})$ can be noted $\tau_k$. We call {\em sequence of footprints} of $\G$ according to $\tau$ the sequence {\em SF}$(\tau) = G^{\tau_0},G^{\tau_1}, \ldots$.

\subsubsection{Expressing other Temporal Concepts}
\label{sec:journeys}

A sequence of couples $\J=\{(e_1,t_1),$ $(e_2,t_2) \dots,$ $(e_k,t_k)\}$, such that $\{e_1, e_2,...,e_k\}$ is a  walk in $G$ is a {\em journey} in $\G$ if and only if $\forall i,  1\leq i < k$, $\rho(e_i,t_i)=1$ and $t_{i+1}\ge t_i$. The $departure(\J)$ and $arrival(\J)$ are respectively the starting date $t_1$ and the last date $t_k$ of a journey $\J$.

Journeys can be thought of as {\em paths over time} from a source to a destination and therefore have both a {\em topological} and a {\em temporal} length.
The {\em topological length} of $\J$ is the number $\topo{|\J|}= k$ of couples in $\J$ (i.e., the number of {\em hops}); its {\em temporal length} is its end-to-end duration:  $\topo{||\J||}= arrival(\J) - departure(\J)$.

\section{Exploring the Dataset}

\subsection{The Dataset}

As mentioned in the introduction, the scientific community analyzed in this work has been extracted from the hep-th (High Energy Physics – Theory) portion of the arXiv website, an on-line repository available at http://arxiv.org/. 

The dataset is composed by a collection of papers and therefore their related citations over the period within January 1992 to May 2003. For each paper the set of authors, the dates of the on-line publications on arXiv.org, and the references are provided.
There are 352 807 citations within the total amount of 29 555 papers written by 59 439 authors. 
The broadness of the time window covered allows us to explore the dataset in order to extract, capture and characterize the evolution of the interactions patterns within the community by means of different data transformations.

\subsection{The Networks Description}

From the dataset, we can easily derive two graphs.  The first, namely the {\em co-authorships network}, having authors as nodes and the undirected links standing for the relation of co-authoring a paper. 
The second, the {\em citations network}, where nodes are the papers and the links (directed) are the references among papers.
More formally, the derived graphs can be defined as:

\begin{itemize}
 \item the \textbf{co-authorship network} as $G_a : (V_a,E)$ where nodes in $V_a$ are the authors and links $e \in E$ connect nodes co-authoring a paper.
\item the \textbf{citations network} as $G_c: (V_c,E)$ where the nodes in $V_c$ are the papers and each edge $e \in E$ corresponds to a reference to another paper.
\end{itemize}

\begin{table}
 \begin{center}
\begin{tabular}[t]{|c|c|c|}
\hline 
Network Indicators & $G_a$ & $G_c$\tabularnewline
\hline
\hline 
Network Diameter & 26 & 37\tabularnewline
\hline 
Network Modularity & 0,706 & 0.617\tabularnewline
\hline 
Network Average Clustering Coefficient & 0.5006 & 0.156\tabularnewline
\hline 
\end{tabular}
\caption{Co-authorships and Citations Graph Measures}
\label{tab:netstat}
\end{center}
\end{table}

In Table \ref{tab:netstat} we provide measures about the citations and collaborations networks.

The diameters - e.g., the longest shortest path between to pairs of nodes (respectively authors and papers) - of both networks have high values. 
The modularity, measuring how a network can be partitioned into modules or subparts, has high values on both graphs. Whereas the average clustering coefficient of the collaborations graph is higher than in the citations graph. 

The networks are composed by several connected islands with few interconnections within them, and the co-authorships network is more clustered of the citations graph.

\section{Temporalizing the Dataset}
 
In this section we are going to explicit the temporal aspects, (i.e. the structural evolution) of the {\em citations} and {\em co-authorships} networks. 
The transformation is performed through the {\em time-varying graphs} formalism defined in the previous section.

We derive two time-varying graphs: the {\em temporal co-authorships network}, with undirected edges and authors as nodes where a link stands for the relations of co-authoring a paper; and the {\em temporal citations network} having papers as nodes and the links (directed) representing the citations from a paper to another one. The temporal dimension of both networks is derived by the paper's submission date. The temporal co-authorship network has edges labeled with the date of submission, while the temporal citations network has the nodes labeled with the publication date of papers citing other papers. 

More formally, we can define

\begin{itemize}
\item the \textbf{temporal co-authorships network} as a quadruplet $G_a^t : (V,E,\T,\rho)$ where
the nodes in $v \in V$ are the authors and links $e \in E$ connect a couple of scientists co-authoring a paper. The temporal domain $\T=[t_a, t_b)$ of the function $\rho$, is the {\em lifetime} of each node $v$ that in this context is assumed as $t_a$ to be the submission date of the paper and $t_b = \infty$.
\item the \textbf{temporal citations network} as a quadruplet $G_c^t: (V,E,\T,\rho)$ where the nodes in the set $V$ are the papers and each edge $e\in E$ corresponds to a citation to another paper. 
As for the co-authorships network, the temporal dimension $\T=[t_a, t_b)$ of the presence function $\rho$ of  $G_c^t$ is defined within the submission date of papers and $\infty$.
\end{itemize}

\subsection{Citations Network Evolution}
\begin{figure}[h!]
 \centering
 \subfigure[Average Clustering Coefficient]
   {\includegraphics[width=55mm]{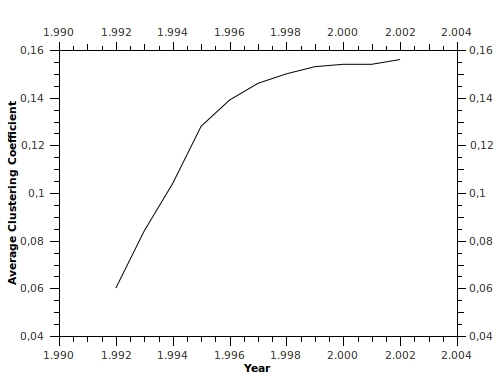} \label{fig:cittemporalclust}}
 \hspace{1mm}
 \subfigure[Density]
 {\includegraphics[width=55mm]{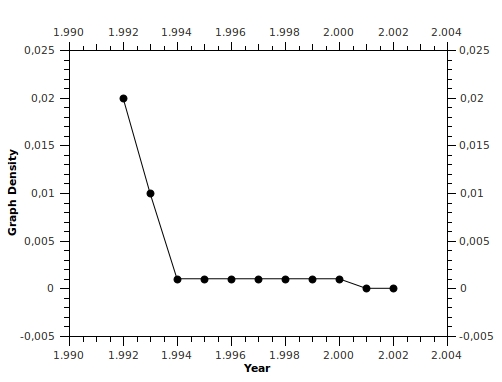} \label{fig:cittemporaldensity}}
\hspace{1mm}
 \subfigure[Modularity]
 {\includegraphics[width=55mm]{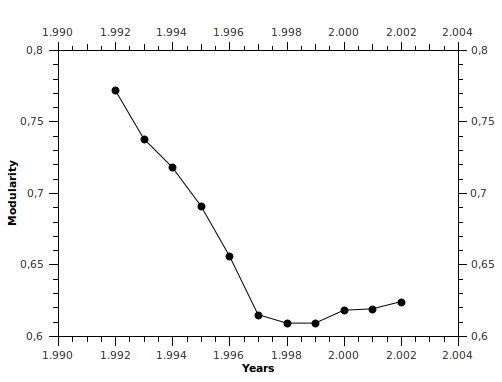} \label{fig:cittemporalmodularity}}
 \caption{Citations Graph Evolution}
\label{fig:cittemporal}
\end{figure}

In this section we show the evolution of the temporal citations network $G_c^t$ by using the sequence of footprints as defined in section \ref{sec:foot}. 
The values are computed by aggregating the interactions occurring at each sub-interval $SF(\tau)$ having $\tau$ fixed to one year.
Figure \ref{fig:cittemporalclust} shows the evolution of the clustering coefficient the curve is characterized by a stable trend attesting on low values. The density evolution, which is shown in Figure \ref{fig:cittemporaldensity}, presents the same low and decreasing behavior, meaning that both the distances and interconnections among nodes (citations within papers) are stable for all the time windows observed. Also the modularity, shown in Figure \ref{fig:cittemporalmodularity}, has a decreasing but stable trend.

\subsection{Co-authorships Network Evolution}

\begin{figure}[h!]
 \centering
 \subfigure[Average Clustering Coefficient]
   {\includegraphics[width=55mm]{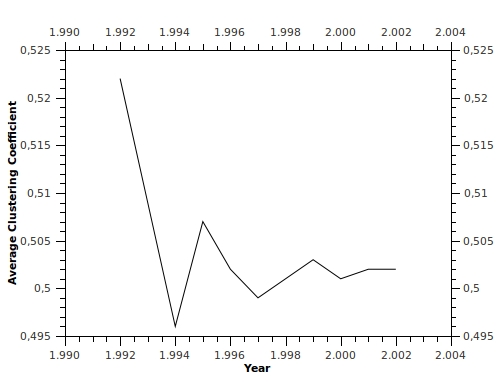} \label{fig:coatemporalclustering}}
 \hspace{1mm}
 \subfigure[Density]
 {\includegraphics[width=55mm]{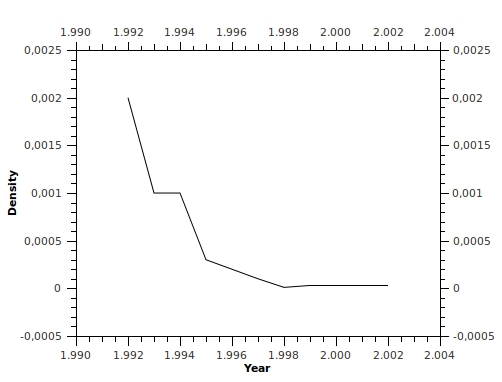} \label{fig:coatemporaldensity}}
\hspace{1mm}
 \subfigure[Modularity]
 {\includegraphics[width=55mm]{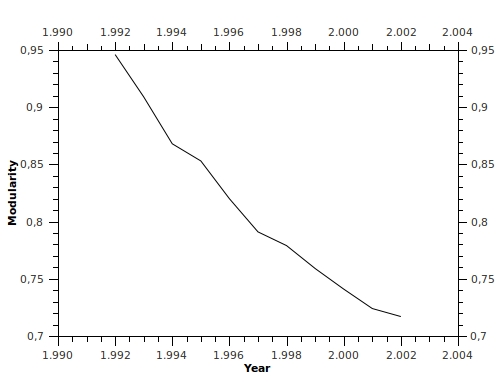} \label{fig:coatemporalmodularity}}
 \caption{Co-authorships Graph Evolution}
\label{fig:coatemporal}
\end{figure}

The temporal co-authorships graphs presents a different structural evolution with respect to the temporal graph of citations analyzed in the previous section. 
As before, here the values are computed by aggregating the interactions at each sub-interval $SF(\tau)$ where  $\tau$ is fixed to one year. 
The average clustering coefficient evolution in the time interval observed, that is shown in Figure \ref{fig:coatemporalclustering}, has an oscillating trend with higher values than the ones reached by the temporal citations graph. In addition, $G_a^t$ has a more modular and denser structure, as shown in Figure \ref{fig:coatemporalmodularity} and Figure \ref{fig:coatemporaldensity}. 
The captured trends are characterized by a decreasing (and not stable) trend.

\section{Expliciting Interactions}

In this section we provide an additional data transformation in order to capture more details about the evolution of our scientific network.
The dynamics in scientific communities are based upon competitions and collaborations among authors and groups of scientists. In the analysis we want to capture a) the resulting emerging effects caused by these two opposite motivations and b) how they are expressed in terms of collaborations and citations patterns. 
The dataset analyzed in the current paper presents two explicit interactions: the papers' co-authorships and the citations between papers.  In addition, there is an implicit level of interaction that depends upon the goals behind any research paper: the quality and, at the same time, the necessity to be highly cited (competition).  Hence, often both the collaborations' and citations' strategies are optimized in order to have the highest impact with respect to the problem addressed and to collect the highest number of citations. How such processes affect the scientific production and the scientific communities structural evolution?

\subsection{Deriving the Interactions Graph}
We approach the data transformation in order to explicit the co-authorships and cited collaboration patterns and their interdependencies. The dataset is transformed in an undirected graph, that we call the {\em interaction network}, having as nodes the authors, weighted links representing the co-authorship on a paper, and when a paper is cited by another work, the links' weights, connecting the authors of the referenced paper, are incremented. 

More formally, the graph of the {\em cited co-authorships} is defined as  quadrupled $G_cc : (V,E,\T,\rho)$  on a discrete time. 
The nodes $v \in V$ are the authors, the set of edges $E$ represents the collaborations on a paper's production. 
The nodes appear on the graph the first time a paper they wrote has been published, and the interaction $L$ is weighted with a variable $w_i$, namely the {\em strength value} of a collaboration, that is incremented at each citation received by a paper produced by a given couple of nodes $(u,v)$. 

In the following section we analyze the behaviors and the interaction's strategies within the \textbf{most cited authors' network}, such a graph, namely $G_i$, is a subset of the global interaction network $G_cc : (V,E,\T,\rho)$. 

In particular the nodes considered in the analysis are only the authors having links' strength values $w_i \geq 150$, that is, all the groups having more than 150 citations on a work. Such a network in its maximum expansion, during the 10 years temporal window observed, is composed by 12 583 nodes and 84 512 edges.

\subsection{The Phase Transition}

\begin{figure}[h!]
 \centering
 \subfigure[Average Clustering Coefficient]
   {\includegraphics[width=50mm]{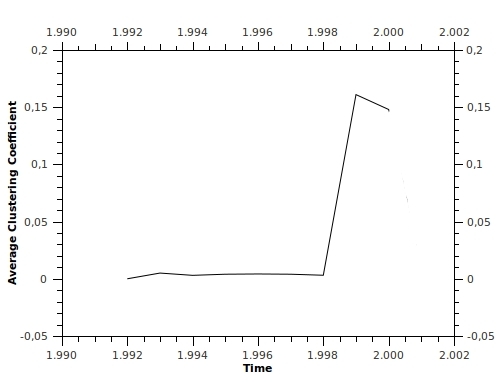} \label{fig:Clustering}}
 \hspace{1mm}
 \subfigure[Density]
 {\includegraphics[width=50mm]{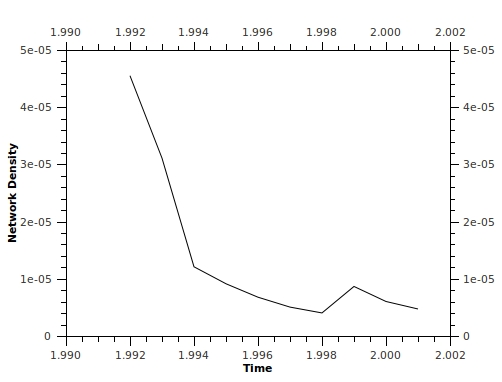} \label{fig:Density}}
\hspace{1mm}
 \subfigure[Modularity]
 {\includegraphics[width=50mm]{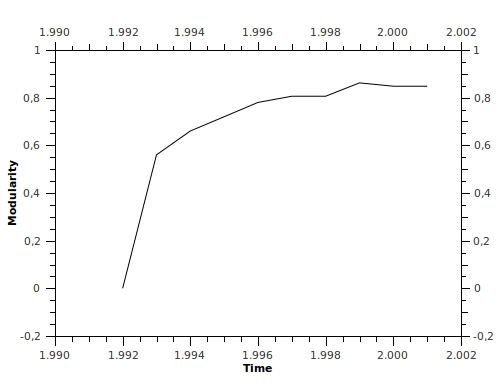} \label{fig:Modularity}}
 \caption{Collaborations Graph Evolution}
\label{fig:interactions}
\end{figure}

Figure \ref{fig:Density} shows the density values for each element of the temporal sequence of footprints $SF(\tau)$ of the interactions network of the most proficient scientists $G_i$. The time interval $\tau$ is fixed to one year.

The density trend starts with very low values and then an increase of the graph's sparsity occurs during its evolution with a very low counter-trend during the period between 1999 and 2000.

 The growing rate of the modularity, computed on $SF(\tau)$ is shown in \ref{fig:Modularity}. It is characterized by an increasing rate until 1993, then it reaches its highest values during the period between 1999 and 2000, but through a smoothed rate. 
As far as we can see by the modularity evolution, the interconnections among separated groups of authors starts in 1993, then their interconnection continues, but with a more stable rate. 
Looking at the curve of the {\em average clustering coefficient} shown in (Figure \ref{fig:Clustering}), we can see a phase transition occurring between 1999 and 2000 and separating a monotone trend from a decreasing one.  

We can interpret the {\em modularity} trend as showing that nodes during the first phase are divided in several and separated groups, while after the phase transition of the clustering coefficient, the connections among these groups start to become denser causing a network structure with a smaller number of larger communities (modules) - e.g. the network tends toward a structural homogeneity.

\subsection{Zooming on Interconnections}

\begin{figure}[h]
 \centering
 \includegraphics[width=70mm]{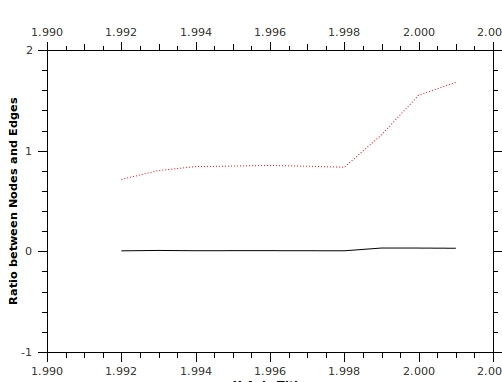}
 \caption{The average number of edges per node for the entire network of the cited co-authorships (in black) and for the network of the most proficient authors (in red)}
 \label{fig:Ratio}
\end{figure}

In order characterize the phenomena behind the phase transition outlined in the previous section, in 
Figure \ref{fig:Ratio} we show the trend of the average ratio between nodes and edges in both the whole interaction network (in black) and the network of the most proficient scientists (in red). 
As we can see, the phase transition, evinced in the clustering coefficient evolution in Figure \ref{fig:Clustering}, is not caused by an increase of the number of authors in the period between 1999 and 2000, neither it is a pattern related to the whole dataset. 

In Table \ref{tab:otherindicators} we present the evolution of the average degree, the average path length and of the degree power law within the temporal window observed. 
As for the previous indicators these values are computed on the sequence of footprints $SF(\tau)$ with $\tau$ fixed to one year of $G_i$.

\begin{table}
\begin{center}
 \begin{tabular}{|c|c|c|c|}
\hline 
Year & Average Degree & Average Path Length & Power Law\tabularnewline
\hline
\hline 
1992 & 0,0095 & 1 & 0\tabularnewline
\hline 
1993 & 0,0176 & 1 & -1,386\tabularnewline
\hline 
1994 & 0,012 & 1 & -1,79\tabularnewline
\hline 
1995 & 0,0135 & 1,16 & -2,16\tabularnewline
\hline 
1996 & 0,132 & 1,13 & -2,27\tabularnewline
\hline 
1997 & 0,0118 & 1,12 & -2,5\tabularnewline
\hline 
1998 & 0,106 & 1,12 & -2,5\tabularnewline
\hline 
\textbf{1999} & \textbf{0,066} & \textbf{3,92} & \textbf{-5,08}\tabularnewline
\hline 
\textbf{2000} & \textbf{0,64} & \textbf{3,79} & \textbf{-5,27}\tabularnewline
\hline 
2001 & 0,6 & 3,82 & -5,25\tabularnewline
\hline
\end{tabular}
\label{tab:otherindicators}
\caption{Other interaction network's measurements}
\end{center}
\end{table}

In bold the values when the phase transition occurs. Neither the average path length, indicating the average distances among nodes, the power law degree, measuring how closely the degree distribution of a network follows a power-law scale and the evolution of average degree, counting the average number of connections at each node, are immune to the phase transition.

\section{Goals and Preferential Attachment}
As the Time-Varying graphs is an interaction-centric formalism. In this section we will show how such a modeling approach is compliant with one of the widely diffused platforms for network analysis and how it is possible to show the punctual evolution as in a movie of the temporal networks.
 
The most interesting emerging phenomenon from the previous section is a phase transition in the evolution of the structure of the most proficient authors' network occurring between 1999 and 2000. According to the temporal analysis, such changes in the network are caused by a particular trend regarding the interconnections among nodes (authors) of the most cited co-authorships graph $G_i$. 

\begin{figure}[h!]
 \centering
 \subfigure[Several separated connected components]
   {\includegraphics[width=45mm]{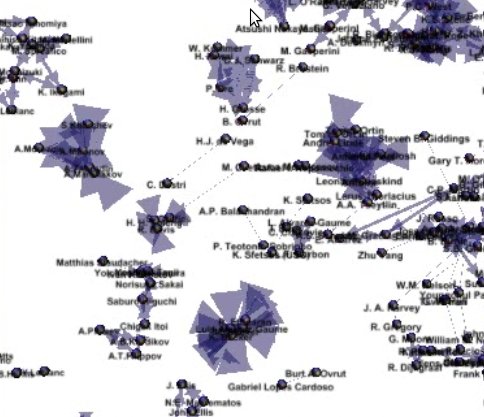} \label{fig:6}}
 \hspace{1mm}
 \subfigure[that start to connect with each other]
 {\includegraphics[width=51mm]{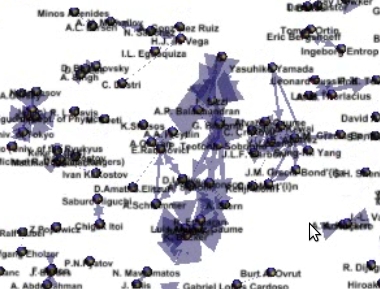} \label{fig:7}}
 \caption{Connections within the islands}
\label{fig:comm1}
\end{figure}

In this section we outline the network evolution by showing the formation of the biggest community at the beginning of the phase transition (1998) until its maximum expansion (2002). In Figure \ref{fig:comm1} we provide a sequence of screen-shots showing the nodes' aggregation patterns. 
The pictures are obtained through the Gephi platform (\cite{Gephi09}). At the beginning (Figure \ref{fig:6}) there are several separated components, that start to connect (Figure \ref{fig:7}).

\begin{figure}[h!]
 \centering
 \subfigure[The number of connections within authors continues to increase.]
   {\includegraphics[width=51mm]{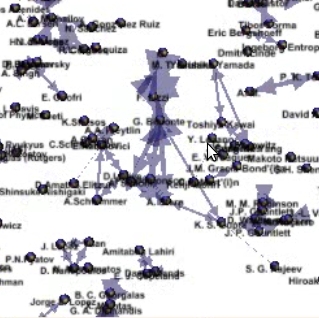} \label{fig:8}}
 \hspace{1mm}
 \subfigure[The maximum level of connectivity is reached.]
 {\includegraphics[width=45mm]{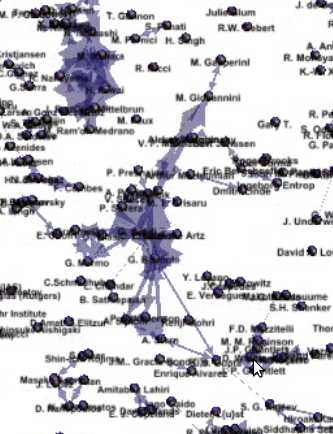} \label{fig:9}}
 \caption{The growing phase of interaction among authors}
\label{fig:comm2}
\end{figure}

Notice that the edges are emphasized with respect to the {\em strength value} $w_i$ counting the number of citations of each couple of nodes. 
The component (group of authors) in the center is highly cited and it is playing as an attractor on the neighboring nodes as it is shown in Figure \ref{fig:8} until the maximum level of connections in the group is reached, as shown in Figure \ref{fig:9}.

\subsection{Zooming on the Attractors}

\begin{figure}[h]
 \centering
 \includegraphics[width=100mm]{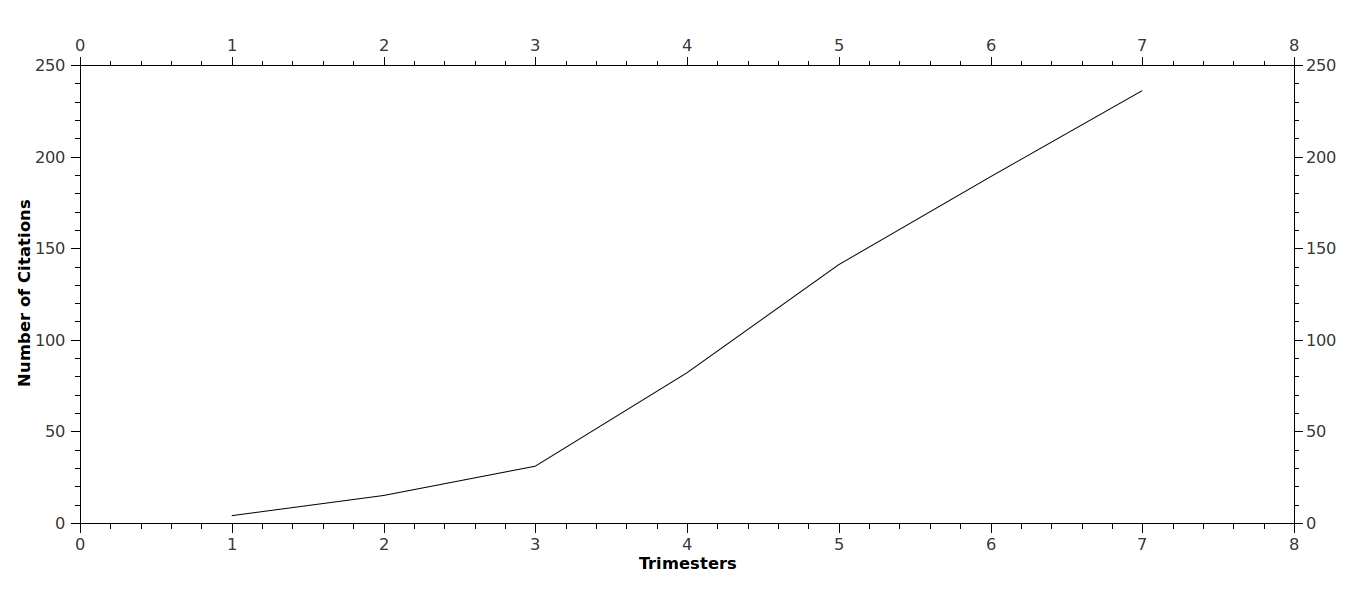}
\caption{the citations trend of the most cited paper for each semester}
\label{fig:bestpapercit}
\end{figure}

In this section we provide a more detailed vision on such a process of aggregation toward the attractors. 
Let starts by introducing Figure \ref{fig:bestpapercit} showing the number of citations received at each semester by the most cited paper in our dataset.

The citations rate has a strong increase after two semesters. The third semester coincides with the interval (1999-2000) of the phase transition captured in the previous section.

Hence, in order to understand the effect of this paper, in the following we will show a sequence of snapshots of the network structure in the neighbor of the authors of the most cited paper when it appears in our database. 
Notice that in the following pictures, the nodes' diameters are proportional to the total amount of citations received by their papers.

\begin{figure}[h!]
 \centering
 \subfigure[Before]
   {\includegraphics[width=60mm]{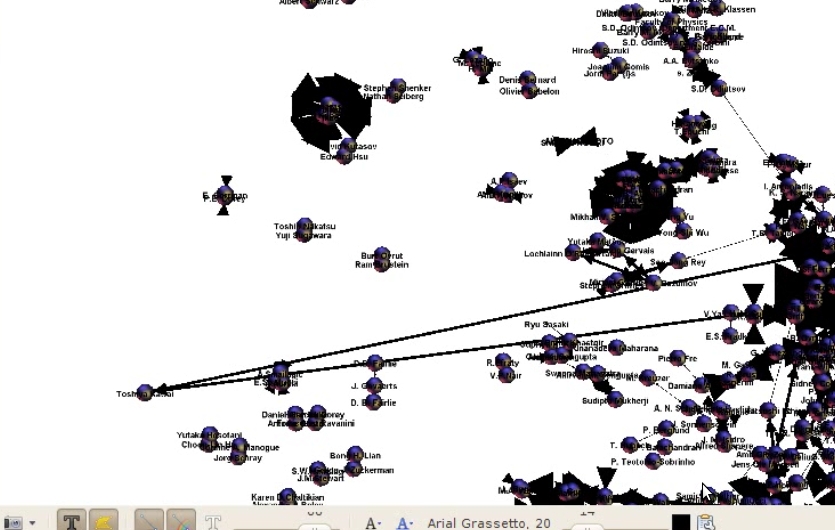} \label{fig:1}}
 \hspace{1mm}
 \subfigure[one of the authors (the biggest node) of the most cited paper and a smaller node with an higher degree appear]
 {\includegraphics[width=50mm]{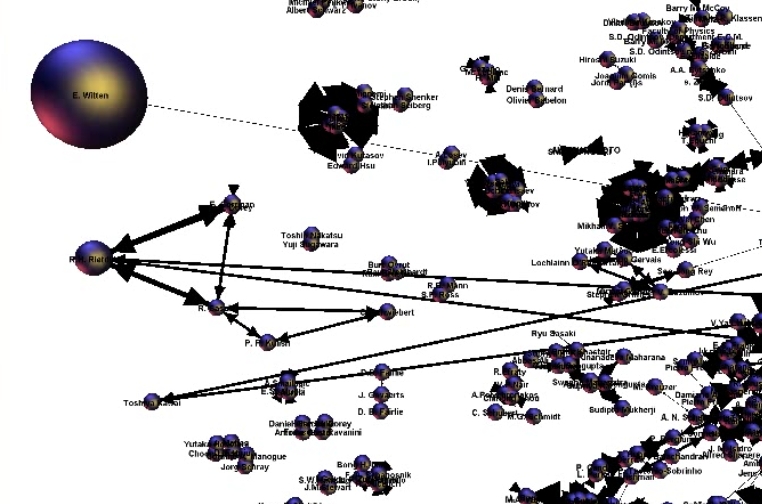} \label{fig:2}}
 \caption{The appearence of one of the most cited authors}
\end{figure}

At the beginning there are only separated components as shown in Figure \ref{fig:1}. Then a large node appears (Figure \ref{fig:2}), and near appears a node with a smaller diameter but with a higher number of links. The biggest node is one of the authors of the most cited paper and as we can see, the node has a very low number of connections (collaborations) in that time interval.

\begin{figure}[h!]
 \centering
 \subfigure[The group of authors of the most cited papers appears. The authors are the two big nodes and the smaller hub]
   {\includegraphics[width=50mm]{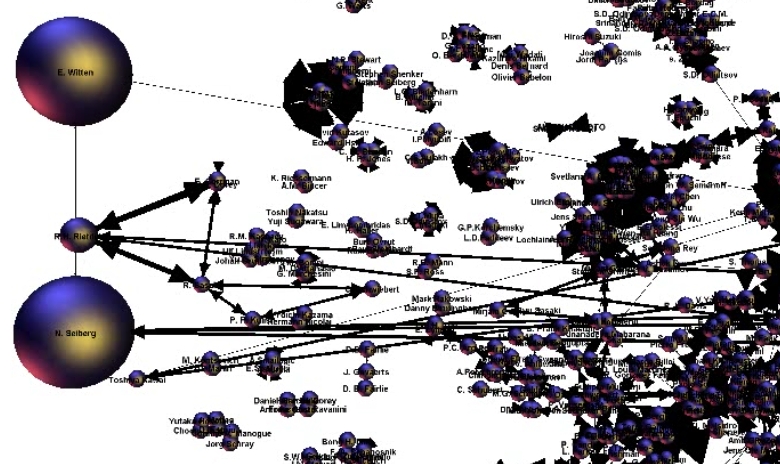} \label{fig:4}}
 \hspace{1mm}
 \subfigure[The portion of the graph becomes denser]
 {\includegraphics[width=50mm]{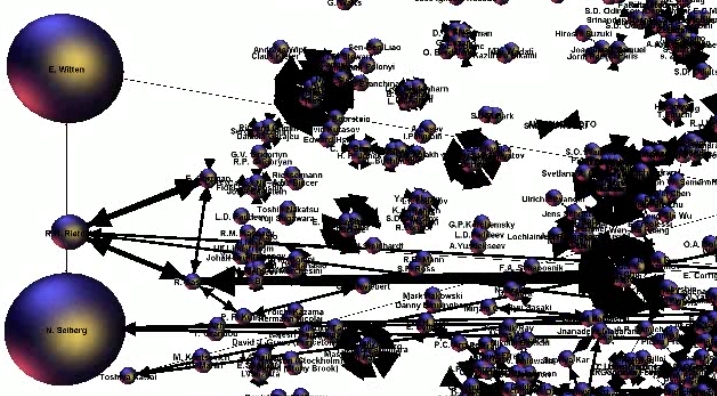} \label{fig:5}}
 \caption{Densification through the hub node}
\end{figure}

In Figure \ref{fig:4} the fat node (a Nobel prize) and the hub node are connected, they publish a paper together with another node with a large diameter. Several islands start to link the clique formed and as we can see in Figure \ref{fig:5} the process of diffusion continues by means of new hubs. 

The sequence of snapshots shows the interactions patterns behind the formulation of the ``String Theory'' and of its consequent developments. 

Authors in the same community start to migrate toward the “island” of the authors of the most cited paper.
The increasing community densification causes the formation of a giant component around the group authoring the most cited paper that in turns makes the network become denser and homogeneous as emerged in the analysis in the previous sections.

It is a goal-driven preferential attachment  – e.g., the mechanism used to explain the power law degree distributions in social networks - due to the number of citations (representing the emergence through selection) to a given group.  Authors tend to join highly cited groups to satisfy both the quality and the possibility to be highly cited requirements.
Moreover, considering that at the beginning there are several separated groups, the phenomenon can be interpreted as a three-fold process with a first phase as the exploration of ideas by means of separated works afforded by separated groups, a second one when a part of the ideas explored starts to be cited more than the others, and a third one when authors tend to join groups that have produced highly cited works. The process tripartition resembles the phases of the natural selection, e.g. the {\em exploration}, the {\em selection} and {\em migration}. In this context  such a (social) selection a) produces self-organization because it is played by a group of individuals which act, compete and collaborate in order to advance science and b) determines the success (emergence) of a topic and of the scientists working on it.

\subsection{Characterizing the Community Evolution}

Table \ref{table:metrics} summarizes the network evolution by means of a) basic indicators, e.g. the number of nodes, the number of edges and the community's diameter) and b) aggregated indicators, e.g. the cyclomatic number, the alpha, beta, and gamma index. 
The {\em cyclomatic number} counts the number of cycles on the graph, its magnitude characterizes the development of the nodes' accessibility. The {\em alpha index} is the ratio between the number of cycles in the graph and their possible maximum value. The range of the alpha index spread within 0 to 1, that are from no cycles to a completely interconnected network.
The {\em beta index}, is a simple measure of connectivity. It relates the total number of edges to the total number of nodes. The higher the value, the greater the connectivity is. The {\em gamma index} measures the ratio between the number of edges on the network and the maximum number of possible edges among nodes. The gamma index spreads within 0 and 100, respectively indicating the minimum and the maximum number of edges between nodes.

As we can see from the evolution of these parameters, the aggregation pattern among separated components is evident for each one of the metric proposed. In terms of nodes that join the community and their mutual connections, the diameter over time passes through a phase of expansion and then tends to stabilize.

\begin{center}
\begin{table}[h!]
 \begin{tabular}{|c|c|c|c|c|c|c|c|}
\hline 
Measures & April 00 & October 00 & April 01 & October 01 & April 02 & October 02 & April 03\tabularnewline
\hline
\hline 
Vertices:             & 23 & 51 & 65 & 66 & 67 & 70 & 72\tabularnewline
\hline 
Edges:                & 29 & 75 & 99 & 100 & 106 & 110 & 114\tabularnewline
\hline 
Diameter:             & 6 & 10 & 10 & 10 & 8 & 8 & 8\tabularnewline
\hline 
Cyclomatic:  & 17 & 25 & 35 & 35 & 40 & 41 & 43\tabularnewline
\hline 
Alpha:  & 0,73 & 0,02 & 0,017 & 0,016 & 0,018 & 0,017 & 0,017\tabularnewline
\hline 
Beta: & 1,69 & 1,47 & 1,52 & 1,51 & 1,58 & 1,57 & 1,58\tabularnewline
\hline 
Gamma: & 61,9 & 51,02 & 52,38 & 52,08 & 54,3 & 53,92 & 54,28\tabularnewline
\hline 
\end{tabular}
\label{table:metrics}
\caption{network measurement of the biggest community}
\end{table}               
\end{center}

\section{Conclusions}
In this paper we characterize the evolution of a scientific community extracted by  the ArXiv's hep-th (High Energy Physics – Theory) repository. 
The analysis starts with a static vision on the dataset by showing the structure of the citations and co-authorships graphs derived by the dataset. Then by adding the temporal dimension on both networks we characterize the structural changes of the co-authorships and citations graphs.
The temporal dimension and the metrics used for the analysis were formalized using Time-Varying Graphs (TVG), a mathematical framework designed to represent the interactions and their evolution in dynamically changing environments. 

Since we are interested in the relationships between collaborations and citations behaviors of scientists, we focus on the network of most cited authors and on its structural evolution where several interesting aspects emerge. The network evolves toward a denser structure, a phase transition occurs in the 1999-2000 time interval causing the homogenization of communities. 

Through our approach, we capture the role played by famous authors on co-authorship behaviors. They act as attractors on the community. The driving force is a sort of preferential attachment driven by the number of citations received by a given group, that in terms of the goal of any scientific community indicates a strategy oriented to the community belonging. 

Furthermore, the evolution of the network from a sparse and modular structure to a denser and homogeneous one, can be interpreted as a three-fold process reflecting the natural selection. The first phase is the exploration of ideas by means of separated works, once some ideas start to be cited (selected) more than others, then authors tend to join groups that have produced highly cited works. 
The selection is performed by individuals in a goal oriented environment and such a (social) selection produces self-organization because it is played by a group of individuals which act, compete and collaborate in order to advance Science. In fact, the driving force is an emergent effect of the interdependencies between citations and the goal of the scientific production since the social selection determines the emergence of a topic and of the scientists working on it by determining the so called preferential attachment toward groups and topics having high potential of citations. Finally, we show that the migration of authors toward the most cited authors (attractors) expresses through a hub node, - e.g. a node with few citations and several co-authorships. 

In the next future we are going to outline the behavior of the most proficient scientist in terms of their aggregation patterns, and on how their works are diffused within the community, that is, characterizing the reasons behind the selection process causing the network structural evolution.
Such aspects will be addressed both with new analyses on different datasets and by means multi-agent simulations. The former stream will be devoted to the definition of new patterns, the latter will be used for the understanding of how changing some parameters of the network influences the evolution, and consequently the quality, of the scientific production.

\section{Acknowledgments}
This work was partially supported by the Future and Emerging Technologies programme FP7-COSI-ICT of the European Commission through project QLectives (grant no.: 231200).

\bibliographystyle{plain}
\bibliography{dtn}

\begin{thebibliography}{10}

\bibitem{Barabasi2002}
Jeong H. Neda Z. Ravasz E. Schubert~A. Barabasi, A.L. and T.~Vicsek.
\newblock Evolution of the social network of scientific collaborations.
\newblock {\em Physica A, vol.311}, pages 590--614, 2002.

\bibitem{Gephi09}
M.~Bastian, S.~Heymann, and M.~Jacomy.
\newblock {Gephi: An open source software for exploring and manipulating
  networks}.
\newblock In {\em International AAAI Conference on Weblogs and Social Media},
  2009.

\bibitem{Carley02}
C.T Butts and K.M. Carley.
\newblock Structural change and homeostasis in organizations: A
  decision-theoretic approach, 2002.

\bibitem{CFQS2010}
A.~Casteigts, P.~Flocchini, W.~Quattrociocchi, and N.~Santoro.
\newblock Time-varying graphs and dynamic networks.
\newblock {\em Technical Report University of Carleton, Canada}, 2010.

\bibitem{castellano07}
C.~Castellano, S.~Fortunato, and V.~Loreto.
\newblock Statistical physics of social dynamics.
\newblock 2007.

\bibitem{Davidsen2002}
J\"orn Davidsen, Holger Ebel, and Stefan Bornholdt.
\newblock Emergence of a small world from local interactions: Modeling
  acquaintance networks.
\newblock {\em Phys. Rev. Lett.}, 88(12):128701, Mar 2002.

\bibitem{amblard01}
G.~Deffuant, D.~Neau, F.~Amblard, and Gerard Weisbuch.
\newblock Mixing beliefs among interacting agents.
\newblock {\em Advances in Complex Systems}, 3:87--98, 2001.

\bibitem{Gilbert77}
N.~Gilbert.
\newblock Referencing as persuasion.
\newblock {\em Social Studies of Science, vol.7}, pages 113--22, 1977.

\bibitem{Guimera05}
R.~Guimera, B.~Uzzi, J.~Spiro, and L.A. Amaral.
\newblock {Team Assembly Mechanisms Determine Collaboration Network Structure
  and Team Performance}.
\newblock {\em Science}, 308(5722):697--702, 2005.

\bibitem{Holme05}
P.~Holme.
\newblock {Network reachability of real-world contact sequences}.
\newblock {\em Physical Review E}, 71(4):46119, 2005.

\bibitem{Hummon89}
Norman~P. Hummon and Patrick Doreian.
\newblock Connectivity in a citation network: The development of dna theory,
  1989.

\bibitem{Klemm02}
Konstantin Klemm and V\'ictor~M. Egu\'iluz.
\newblock {Highly clustered scale-free networks}.
\newblock {\em Physical Review E}, 65(3):036123+, Feb 2002.

\bibitem{KosKW08}
G.~Kossinets, J.~Kleinberg, and D.~Watts.
\newblock {The structure of information pathways in a social communication
  network}.
\newblock In {\em Proc. of the 14th ACM SIGKDD Intl. Conf. on Knowledge
  Discovery and Data Mining (KDD 2008)}, pages 435--443, 2008.

\bibitem{Kostakos09}
V.~Kostakos.
\newblock {Temporal graphs}.
\newblock {\em Physica A: Statistical Mechanics and its Applications},
  388(6):1007--1023, 2009.

\bibitem{Lelarge09}
M.~Lelarge.
\newblock Diffusion of innovations on random networks: Understanding the chasm,
  2009.

\bibitem{Les05b}
Jure Leskovec, Deepayan Chakrabarti, Jon Kleinberg, and Christos Faloutsos.

\bibitem{LESK10}
Jure Leskovec, Deepayan Chakrabarti, Jon~M. Kleinberg, Christos Faloutsos, and
  Zoubin Ghahramani.
\newblock Kronecker graphs: An approach to modeling networks.
\newblock {\em Journal of Machine Learning Research}, 11:985--1042, 2010.

\bibitem{Les05}
Jure Leskovec, Jon Kleinberg, and Christos Faloutsos.

\bibitem{LES07}
Jure Leskovec, Jon~M. Kleinberg, and Christos Faloutsos.
\newblock Graph evolution: Densification and shrinking diameters.
\newblock {\em TKDD}, 1(1), 2007.

\bibitem{MacRoberts96}
M.H. MacRoberts and B.R. MacRoberts.
\newblock Problems of citation analysis.
\newblock {\em Scientometrics, vol.36, no.3}, pages 435--444, 1996.

\bibitem{Mataric92}
M.~Mataric.
\newblock {Designing emergent behaviors: From local interactions to collective
  intelligence}.
\newblock In {\em In Proceedings of the International Conference on Simulation
  of Adaptive Behavior: From Animals to Animats}, volume~2, pages 432--441,
  1992.

\bibitem{Moore2000}
C.~Moore and M.E.J. Newman.
\newblock Epidemics and percolation in small-world networks.
\newblock {\em Phys. Rev. E}, 61:5678--5682, 2000.

\bibitem{Newman2001a}
M.~E.~J. Newman.
\newblock Clustering and preferential attachment in growing networks.
\newblock In {\em Physical Review E, vol.64}, 2001.

\bibitem{Newman2001}
M.~E.~J. Newman.
\newblock {The structure of scientific collaboration networks}.
\newblock 98(2):404--409, January 2001.

\bibitem{Newman2004}
M.~E.~J. Newman.
\newblock Coauthorship networks and patterns of scientific collaboration.
\newblock In {\em Proceedings of the National Academy of Sciences}, pages
  5200--5205, 2004.

\bibitem{Newman2004a}
M.~E.~J. Newman.
\newblock Who is the best connected scientist? a study of scientific
  coauthorship networks.
\newblock In {\em Complex Networks, lecture notes in Physics}, 2004.

\bibitem{peltomaki2006}
Matti Peltomaki and Mikko Alava.
\newblock Correlations in bipartite collaboration networks.
\newblock {\em J.STAT.MECH.}, page P01010, 2006.

\bibitem{Powell05}
W.W Powell, D.R. White, and K.W. Koput.
\newblock Network dynamics and field evolution: The growth of
  interorganizational collaboration in the life sciences.
\newblock {\em American Journal of Sociology, vol.110, no.4}, pages 1132--1205,
  2005.

\bibitem{Solla1965}
D.J. De~Solla Price.
\newblock Networks of scientific papers.
\newblock {\em Science, vol.149, no.3683}, pages 510--515, 1965.

\bibitem{quattrociocchi2010e}
W.~Quattrociocchi, R.~Conte, and E.~Lodi.
\newblock Simulating opinion dynamics in heterogeneous communication systems.
\newblock {\em ECCS 2010 - Lisbon Portugal}, 2010.

\bibitem{QuattrociocchiPC09}
W.~Quattrociocchi, M.~Paolucci, and R.~Conte.
\newblock On the effects of informational cheating on social evaluations: image
  and reputation through gossip.
\newblock {\em International Journal of Knowledge and Learning},
  5(5/6):457--471, 2009.

\bibitem{QA2010a}
Walter Quattrociocchi and Frederic Amblard.
\newblock Selection in scientific networks.
\newblock {\em arxiv:1012.4396}, 2010.

\bibitem{Radicchi2009}
F~.Radicchi, S.~Fortunato, B.~Markiness, and A.~Vespignani.
\newblock Diffusion of scientific credits and the ranking of scientists.
\newblock {\em Physical Review E, vol.80}, 2009.

\bibitem{Redner05}
S.~Redner.
\newblock Citation statistics from 110 years of physical review.
\newblock {\em Physical Review, Physics Today, vol.58}, pages 49--54, 2005.

\bibitem{ACFQS2010a}
N.~Santoro, W.~Quattrociocchi, P.~Flocchini, A.~Casteigts, and F.~Amblard.
\newblock Time varying graphs and social network analysis: Temporal indicators
  and metrics.
\newblock {\em Technical Report University of Carleton, Canada}, 2010.

\bibitem{TSM+09}
J.~Tang, S.~Scellato, M.~Musolesi, C.~Mascolo, and V.~Latora.
\newblock {Small-world behavior in time-varying graphs}.
\newblock {\em Arxiv preprint arXiv:0909.1712}, 2009.

\bibitem{Roth10b}
Carla Taramasco, Jean-Philippe Cointet, and Camille Roth.
\newblock Academic team formation as evolving hypergraphs.
\newblock {\em Scientometrics}, April 2010.

\bibitem{Wagner2005}
C.S Wagner and K.~Leydesdorff.
\newblock Network structure, self-organization, and the growth of international
  collaboration in science.
\newblock {\em Research Policy vol 34 n10}, pages 1608--1618, 2005.

\bibitem{WATTS99}
Duncan~J. Watts.
\newblock Networks, dynamics and the small world phenomenon.
\newblock {\em AJS}, 1999.

\bibitem{Woolley1994}
David~R. Woolley.
\newblock {PLATO: The Emergence of Online Community}, 1994.

\end{thebibliography}

\printindex
\end{document}